\documentclass[a4,useAMS,usenatbib,usegraphicx]{mn2e}
\def\apss{\ref@jnl{Ap\&SS}} 
\def\aapr{\ref@jnl{A\&A~Rev.}}
\def\na{\ref@jnl{New~Astronomy}}

\usepackage{amsmath}
\include{epsf}
\usepackage{float}
\usepackage{color}
\usepackage{multirow}

\usepackage{epsf}
\epsfclipon
\title[Ram Pressure drag]{Ram pressure drag - the effects of ram pressure on dark matter and stellar disk dynamics}
\author[R.Smith et al]{R.Smith$^{1}$\thanks{E-mail:rsmith@astro-udec.cl}, M. Fellhauer${^1}$, P. Assmann${^1}$ \\
$^{1}$Departamento de Astronomia, Universidad de Concepcion, Casilla 160-C, Concepcion, Chile}
\begin{document}

\date{Accepted October 25 2011}

\pagerange{\pageref{firstpage}--\pageref{lastpage}} \pubyear{2011}

\maketitle

\label{firstpage}

\begin{abstract}
We investigate the effects of ram pressure stripping on gas-rich disk galaxies in the cluster environment. Ram pressure stripping principally effects the atomic gas in disk galaxies, stripping away outer disk gas to a truncation radius. We demonstrate that the drag force exerted on truncated gas disks is passed to the stellar disk, and surrounding dark matter through their mutual gravity. Using a toy model of ram pressure stripping, we show that this can drag a stellar disk and dark matter cusp off centre within it's dark matter halo by several kiloparsecs. We present a simple analytical description of this process that predicts the drag force strength and its dependency on ram pressures and disk galaxy properties to first order. The motion of the disk can result in temporary deformation of the stellar disk. However we demonstrate that the key source of stellar disk heating is the removal of the gas potential from within the disk. This can result in disk thickening by approximately a factor of two in gas-rich disks.
\end{abstract}


\section{Introduction}
Dynamical studies of the Virgo cluster reveal that it is far from a relaxed structure, consisting of numerous sub-clouds apparently in the process of merging with each other (\citealp{Gavazzi1999}). As time evolves, clusters like Virgo grow in mass by the accretion of field galaxies and galaxy groups. Freshly accreted late-type disk galaxies face new environmental mechanisms, particular to the cluster environment, that can influence their evolution. The presence of a low density, high temperature ionised gas, clearly visible in X-ray observations (\citealp{Matsumoto2000}), forms the basis of the intra-cluster medium (ICM). Motion of a galaxy through this medium is thought to cause ram pressure stripping of a galaxy's atomic gas (\citealp{GunnGott1972}).

Stripping of atomic gas begins in a galaxy's outer disk, migrates radially inwards, and only halts at a radius where the galaxy's disk self-gravity is sufficient to maintain the disk gas against the ram pressure. This radius is often referred to as the `truncation radius'. Observations of giant disk galaxies in the centre of the Virgo cluster support this notion, with central disk galaxies displaying truncated atomic gas (HI) disks (\citealp{Chung2008}). Indeed, the degree by which the disk is truncated is well correlated with the HI deficiency of the disk (\citealp{Boselli2006}) as would be expected in the ram pressure stripping scenario.

In general, the influence of the intra-cluster medium is
considered to primarily effect the diffuse, atomic gas component of late-type disk galaxies. The HI gas clouds have a large
enough cross-section to the ICM wind to be significantly
effected by the ram pressure. Meanwhile, the cross
sectional area of H$_2$ clouds (Quilis et al. 2000) and stars are
simply too small to feel any signiﬁcant ram pressure directly
from the ICM. The loss of the outer disk gas appears to halt star formation at these radii - \cite{Koopmann2006} show that galaxies that are deficient in HI, typically have smaller H$_\alpha$ disks. However the influence of ram pressure stripping on the stellar dynamics of a disk galaxy is little discussed in the literature.

The HI gas component of a disk galaxy must contribute to some degree to the potential of the disk. Therefore it's removal by ram pressure stripping might be expected to have some influence on the stellar dynamics of the remaining stellar disk. However, typically the mass of a giant galaxy's gas disk is a small fraction of their total disk mass, so stripping the gas is not expected to strongly disturb the stellar disk.
 
In fact there have been few studies of the consequential influence of the ram pressure
force on the other components of a galaxy. In the early ram
pressure stripping simulations of \cite{Farouki1980},
the effect of removal of outer disk gas on the stellar disk
was considered. Here, the removal of the gas component from the outer disks of giant spirals resulted in mild
thickening of the stellar disk as a result of loss of disk potential. The effects were limited to the outer disk as the
surface density of these massive disk galaxies prevented gas
removal in their inner regions. Such mass-loss might be expected to cause disk expansion out of the plane of the disk
from both disk faces. This effect might be expected
to be stronger in lower mass late-type galaxies, who tend to
contain much higher gas fractions within their disk. \cite{Lisker2007} also comments that `signiﬁcant mass-loss due
to stripping might effect the stellar configuration of galaxies'. 

\cite{Schulz2001} reported a
more uni-directional effect on the stellar disks of their model galaxies - the stellar and gas disk are reported to be
dragged ∼2 kpc in the direction of the wind. They suggest that the off-centre position of their disk
places the gas disk under a compressive force, trapped between ram pressure on one side, and the restoring force of
the halo. This compressive force can result in the formation
of flocculent spiral structure, and hence angular momentum
loss, the net result of which can raise the gas density in the inner disk,
`annealing' the galaxy to further stripping. \cite{Schulz2001} focus primarily on the impact on the gas disk and little emphasis is placed on the impact on the dark matter component of their galaxy models. 

We extend this study further, emphasising that not only the stellar disk is
dragged, but additionally the central dark matter surrounding the disk also. \cite{Fujita2006} note that if the gas disk is dragged, it is inevitable that this will be communicated to all components of the galaxy, by virtue of their mutual gravity. 

\cite{Schulz2001} state that the stellar component of their models appears largely unaffected by being dragged off-centre. However we see evidence that the reduction in total disk mass by loss of the gas component results in disk thickening in our models. Furthermore,
the motion of the disk and central dark matter can temporarily disturb
the stellar disk configuration causing a brief cone-like morphology.

We demonstrate that ram pressure dragging can result in significant displacement of the gas and stellar disk, and central dark matter during the infall of a galaxy into a cluster. We provide a simple analytical formulation of the force provided by ram pressure dragging. We demonstrate that it predicts the drag force dependency on disk galaxy properties and ram pressures to first order by comparison with wind-tunnel tests. 

We describe the numerical code, disk galaxy models, and ram pressure model in Section 2, we present our results for a galaxy infall in Section 3.1, the analytical description and testing by wind-tunnel tests in Section 3.2, and impact on stellar disk heating in Section 3.3, our discussion is in Section 4, and we draw conclusions in Section 5.

\section{Setup}
\subsection{Disk galaxy models}
Our dwarf galaxy models consist of 3 components; an NFW dark matter halo (\citealp*{Navarro1996}), an exponential disk of gas and one of stars. The methods used to form each component will be discussed briefly in the following sections. For a more in-depth description please see \cite{Smith2010a}.
\subsubsection{The dark matter halo}
The dark matter halos of all dwarf galaxy models presented in this work have an NFW density profile. The NFW profile has the form:
\begin{equation}
\rho(R) = \frac{\rho_0}{(\frac{r}{r_{\rm{s}}})(1+\frac{r}{r_{\rm{s}}})^2}
\label{NFWdensprof}
\end{equation}
\noindent where $r_{\rm{s}}$ is a characteristic radial scale-length. The profile is truncated at the Virial radius, $r_{\rm{200}}=r_{\rm{s}} c$. Here $c$ is the concentration  parameter (\citealp{Lokas2001}). $c$ is found to have a range of values in cosmological simulations, however there is a general trend for higher values in less massive systems with some scatter - see Figure 8 in \cite{Navarro1996}.

Positions and velocities are assigned to the dark matter particles using the publically available algorithm {\it{mkhalo}} from the {\sc{nemo}} repository (\citealp{McMillan2007b}). Dark matter halos produced in this manner are evolved in isolation for 2.5 Gyr to test stability, and are found to be highly stable.

\subsubsection{The galaxy disks}
The stellar and gas disk both have an exponential form
\begin{equation}
\label{expdisk}
\Sigma(R) = \Sigma_0 {\rm{exp}} (R/R_{\rm{d}})
\end{equation}

\noindent
where $\Sigma$ is the surface density, $\Sigma_{\rm{0}}$ is central surface density, $R$ is radius within the disk, and $R_{\rm{d}}$ is the scale-length of the disk.

The size of the stellar scale-length of the disk is chosen following the recipe of \cite{Mo1998}. Here, the disk mass is a fixed fraction, $m_{\rm{d}}$ of the halo mass. Additionally we must choose the spin parameter $\lambda$. In the Mo et al. recipe, this choice of parameters fully defines the scale-length of the stellar disk, $R_{\rm{d}}$. For simplicity we assume the scale-length of the stellar disk and gas disk are equal.

Disk particles are initially laid down in a plane in their exponential disk following Eq. \ref{expdisk}. Real disks have finite thickness, but we cannot solve for this until we have defined the velocity dispersion throughout the disk. A radially varying velocity dispersion is chosen that ensures the disk is Toomre stable (\citealp{Toomre1964}) at all radii . The Toomre stability criterion is defined as
\begin{equation}
\label{Toomreeqn}
Q \equiv \frac{\kappa\sigma_{R}}{3.36 G \Sigma} > 1
\end{equation}

\noindent
where $\Sigma$ is the surface density, $\sigma_{\rm{R}}$ is the radial velocity dispersion, and $\kappa$ is the epicyclic frequency defined, using the epicyclic approximation (\citealp{SpringelWhite1999}). Next we use $\sigma_{\rm{\phi}}^2 = \frac{\sigma_{\rm{R}}^2}{\gamma^2}$ where

\begin{equation}
\gamma^2 \equiv \frac{4}{\kappa^2 R} \frac{d\Phi}{dR} {\rm{,}}
\end{equation}
\noindent
and $\phi_{\rm{z}} = 0.6 \cdot \phi_{\rm{R}}$ (\cite{Shlosman1993}). This completely defines the minimum values of the necessary velocity dispersions throughout the disk. In practice, $Q>1.5$ is required throughout the stellar disk to ensure stability. The gas disk has an intrinsic velocity dispersion, due to its isothermal nature, that automatically satisfies the Toomre criteria at all radii. 

Once more following \cite{SpringelWhite1999}, the vertical scale height of the disk is defined as
\begin{equation}
z_{\rm{d}} = \frac{\sigma_{\rm{R}}^2}{\pi G \Sigma}{\rm{.}}
\end{equation}

We distribute the particles vertically out of the disk following Spitzer's isothermal sheet solution defined as

\begin{equation}
\rho(R,z) = \frac{\Sigma(R)}{2z_{\rm{d}}} {\rm sech}\,^2(z/z_{\rm{d}}){\rm{.}}
\end{equation}

Finally, the circular velocities of disk particles are calculated. The value of the potential is calculated in thin spherical shells for the combined dark matter, stars and gas in a one-off N-body calculation, including the effects of gravitational softening. This is necessary as the disks own self-gravity can influence the radial orbits of disk particles close to the centre of the halo. The gradient of the potential is then used to calculate the circular velocity at each radius. As a result inner disk particles have circular velocities raised beyond that of the circular velocity of the halo alone at that radius.

\subsubsection{Summary of disk galaxy parameters}
\begin{figure}
  \centering \epsfxsize=8.5cm \epsfysize=6.5cm \epsffile{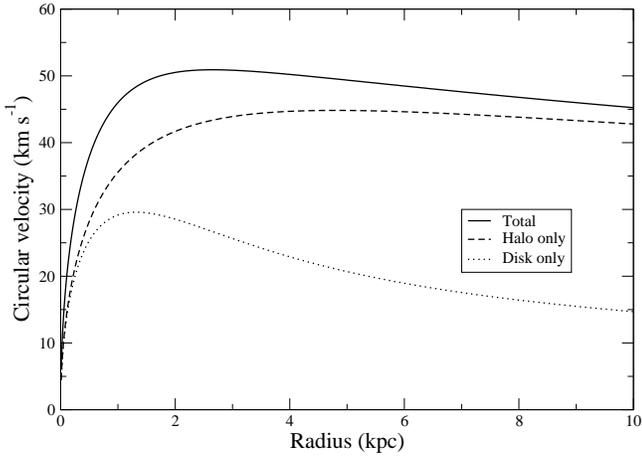}
  \caption{The rotation curve of the galaxy (solid curve) with contributions from the halo (dashed curve) and disk (dotted curve)}
\label{rotcurve}
\end{figure}

The values of parameters defining our standard disk galaxy model can be summarised in the following;
In all simulations our disk galaxy model has a dark matter halo mass of $10^{10}$M$_\odot$, consisting of 100,000 dark matter particles, with a concentration $c=20$. Our standard model has $m_{\rm{d}}$=0.05, therefore has a total disk mass of $5.0 \times 10^{8}$M$_\odot$, and is formed from 50,000 equal mass particles. The stellar to gas ratio of the disk is unity unless otherwise specified. This creates a gas-rich disk galaxy, although such a high gas fractions are not uncommon in low mass disk galaxies (\citealp{Gavazzi2008}). We choose the spin parameter of our standard model as $\lambda$=0.05 resulting in a stellar and gas disk scale length of $R_{\rm{d}}=0.73$ kpc. The rotation curve of this galaxy is presented in Fig. \ref{rotcurve}.

\subsection{The Code}
In this study we make use of `gf' (\citealp{Williams2001},\citealp{Williams1998}), which is a Tree code-SPH algorithm that operates primarily using the techniques described in \cite{Hernquist1989}. While the Tree code allows for rapid calculation of gravitational accelerations, the SPH code allows us to include a HI gas component to our dwarf galaxy models. In all simulations, the gravitational softening length, $\epsilon$, is fixed for all particles at a value of 100 pc, in common with the harassment simulations of \cite{Mastropietro2005}. Gravitational accelerations are evaluated to quadrupole order, using an opening angle $\theta_{\rm{c}}$=0.7. A second order individual particle time-step scheme was utilised to improve efficiency following the methodology of \cite{Hernquist1989}. Each particle was assigned a time-step that is a power of two division of the simulation block time-step, with a minimum time-step of $\sim$5.0 yr. Assignment of time-steps for collisionless particles is controlled by the criteria of \cite{Katz1991}, whereas SPH particle time-steps are assigned using the minimum of the gravitational time-step and the SPH Courant conditions with a Courant constant, $C$=0.1 (\citealp{Hernquist1989}). As discussed in \cite{Williams2004}, the kernel radius $h$ of each SPH particle was allowed to vary such that at all times it maintains between 30 and 40 neighbours within 2$h$. In order to realistically simulate shocks within the SPH model, the artificial viscosity prescription of \cite{Gingold1983} is used with viscosity parameters $(\alpha,\beta)$ = (1,2). The equation of state for the gas component of the galaxies is isothermal with a bulk velocity dispersion of 8.0 km\,s$^{-1}$, in agreement with the measured velocity dispersion of molecular clouds in the local interstellar medium (\citealp{Stark1989}), and the observed HI velocity dispersion within a radius containing significant star formation in late-type disks (\citealp{Tamburro2009}). By choosing an isothermal equation of state, we are intrinsically assuming that stellar feedback processes are balanced by radiative cooling producing a constant velocity dispersion. Such a model can be considered a first approximation to a detailed description of a multiphase ISM that is unresolved in our models, and has been used previously in simulations of galaxy dynamics and galaxy mergers (\citealp{Theis1993}, \citealp{Mihos1994II}, \citealp{Englmaier1997}). We do not include star formation in these models.

\subsection{The Ram Pressure Model}
The Ram Pressure Stripping model is largely identical to that presented in \cite{Vollmer2001}. In this model, additional acceleration vectors are added to individual gas particles to mimic the ram pressure. A live ICM component is not included.
For an individual gas cloud, moving through the ICM of density $\rho_{ICM}$, with a velocity $v$, the pressure on its surface due to sweeping through the medium is assumed to be
 
\begin{equation}
\label{RPSpress}
P_{\rm{ram}} = \rho_{\rm{ICM}} v^2
\end{equation}

In order to calculate strength of the acceleration that gas clouds will feel as a result of ram pressure we follow \cite{Vollmer2001}. A constant column density is assumed for each individual cloud. This has the advantage that the acceleration due to ram pressure is the same for all clouds, disregarding their masses. A value of $\Sigma_{\rm{cld}} = 7.5 \times 10^{20}$ cm$^{-2}$ is used. This is comparable with measurements made by \cite{Rots1990} and \cite{Crosthwaite2000} on nearby face-on galaxies. The acceleration due to ram pressure can therefore be written as

\begin{equation}
\label{RPSforce}
a_{\rm{ram}} = \frac{P_{\rm{ram}}}{m_{\rm{H}}\Sigma_{\rm{cld}}}
\end{equation}
\noindent
where $m_{\rm{H}}$ is the mass of a hydrogen atom. In the galaxy model's frame of reference, the effect of its motion through the ICM is that of a ICM wind of velocity $v$ and density $\rho_{\rm{ICM}}$. $a_{\rm{ram}}$ therefore always acts in the direction of the velocity vector of the wind.

Once more following \cite{Vollmer2001}, a shading criteria is used to select gas particles that feel the influence of ram pressure, and gas particles that are shielded by other gas particles upstream in the wind. In the simulation, gas particles are point particles and therefore have no cross-section by which to shield other particles. However, we can calculate a particles cloud radius $r_{\rm{cld}}$ if we know its mass $m_{\rm{p}}$ and once more assume it has the column density $\Sigma_{\rm{cld}}$. Then $r_{\rm{cld}}$= $\left[m_{\rm{p}}/(\pi  m_{\rm{H}} \Sigma_{\rm{cld}})\right]^{0.5}$. In practice, each particle extends an imaginary vector along the direction of motion of the galaxy's disk and checks to see if any other particles cross-sections ($\pi r_{\rm{cld}}^2$) cross the vector. If there are none, then the gas particle is unshielded and will feel the acceleration $a_{\rm{ram}}$. In this case, an additional acceleration vector of magnitude $a_{\rm{ram}}$ is added to the particle's equation of motion, in the direction of the ICM wind.

We emphasise that this model should be regarded as a toy model of ram pressure. Recent ram pressure stripping models discussed in the literature (see for example \citealp{Roediger2007}) have advanced significantly beyond the simple ram pressure recipe of \cite{Vollmer2001}. Increasingly high resolution along the ISM-ICM boundary has allowed quantification of additional stripping mechanisms such as Rayleigh-Taylor (RT) instabilities and Kelvin-Helmholtz (KH) stripping. The ICM gas is normally composed of a live gas component that can form a shock front when a galaxy reaches supersonic velocities. Typically these studies concentrate on highest possible resolution of the gas component of their disk galaxies, using analytical static potentials to treat the gravitational influence of the dark matter halo and stellar disk. 

Our toy model does not include a live ICM gas component, so physically does not include the effects of RT instabilities or KH stripping. However, the toy model presented is fast, enabling wider parameter searches to be conducted and allowing us to include a live dark matter halo and live stellar disk component in our galaxy models. We will show that this is crucial to the results of this study. We demonstrate in the following section that despite it's simplicity, the toy model can reasonably reproduce the evolution of the HI disk truncation radius in the same manner seen in significantly more complex ram pressure simulations. This requirement is important to the conclusions of our study.

For reasons that will become clear in Section \ref{results}, we clarify that additional accelerations due to ram pressure are applied purely to the gas component of our galaxy models.

\subsubsection{Testing the ram pressure model}
\label{gandgtests}
Despite the large variety of ram pressure stripping models previously implemented in the literature, it has frequently been found that the Gunn $\&$ Gott condition (Eq. \ref{GandGeqn}) can successfully predict the truncation radius $R_{\rm{trc}}$ (the radius to which the gas disk is stripped) to first order as long as the disk is not close to an edge-on inclination to the wind (\citealp{Abadi1999}, \citealp{Vollmer2002}, \citealp{Mayer2005}, \citealp{Jachym2007}, \citealp{Roediger2007}). Within $R_{\rm{trc}}$, the self-gravity of the disk can overcome the ram pressure force.

The Gunn $\&$ Gott condition is 

\begin{equation}
\label{GandGeqn}
\rho_{\rm{ICM}}v^2 = \geq 2\pi G \sigma_{\rm{\star}}(r) \sigma_{\rm{g}}(r){\rm{,}}
\end{equation}
\noindent
where  $v$ is the galaxy's velocity through the ICM, ${\sigma}_{\rm{g}}$ is the gas surface density of the galaxy, and $\sigma_{\rm{\star}}$ is the stellar surface density of the galaxy. The left hand side of Eq. \ref{GandGeqn} represents the removal force due to ram pressure stripping. The right hand side represents the restoring force of the disk's gravitational potential. Hence when the disk is close to face-on inclination to the wind, the self-restoring force of the disk has vectors in the opposite direction from the removal force of ram pressure stripping, and the Gunn $\&$ Gott criteria is best satisfied.

\cite{Abadi1999} showed that the halo and bulge components of a galaxy may provide additional restoring force on the gas disk of the galaxy. For the halo this was found to be negligible, except in the outer disk. However, for galaxies with prominent bulges, the restoring force can be significantly enhanced in the central regions of disk. For our bulgeless disk galaxy models, we will demonstrate that Eq. \ref{GandGeqn} is sufficient.

Using Eq. \ref{GandGeqn}, the predicted truncation radius for a bulgeless galaxy with an exponential disk can be written:

\begin{equation}
\label{truncrad}
R_{\rm{trc}}=-\left(\frac{1}{R_\star}+\frac{1}{R_{\rm{g}}}\right)^{-1} {\rm{ln}}\left[\frac{2 \pi \rho_{\rm{ICM}}v^2 R_{\rm{g}}^2 R_{\rm{\star}}^2}{G m_{\rm{\star}} m_{\rm{g}}}\right]
\end{equation}
\noindent
where $R_\star$ and $R_{\rm{g}}$ are scalelengths of the stellar and gas disk respectively, and $m_{\rm{\star}}$ and $m_{\rm{g}}$ are the total masses of the stellar and gas disk respectively.

\begin{figure}
  \centering \epsfxsize=8.5cm \epsfysize=6.5cm \epsffile{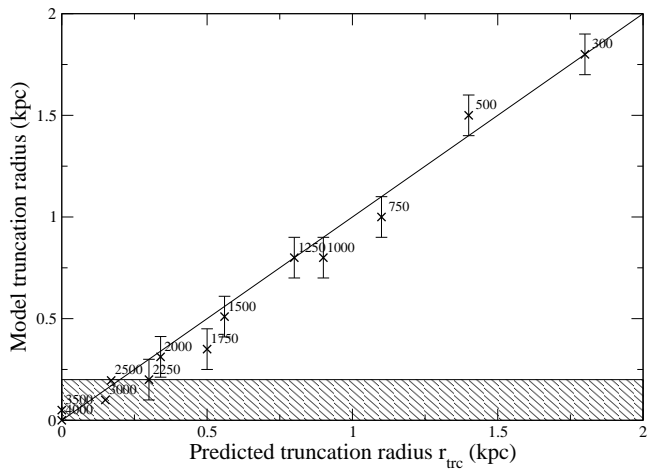}
  \caption{Comparison between the analytically predicted gas disk truncation radius using the Gunn \& Gott Equation (x-axis), and the measured gas disk truncation radius of the model (y-axis). The dashed line marks a one-to-one agreement. In general the agreement is reasonable. The numbers labelling each point represent the ICM wind velocity $v$ used for that wind tunnel test. The hatched area marks the region in which gravity is not fully resolved in the model.}
\label{gandgcomp}
\end{figure}

We use the predicted $R_{\rm{trc}}$ of the disk gas from Eq. \ref{truncrad} to test our ram pressure model in face-on ram pressure stripping simulations. This is more easily accomplished by fixing a constant velocity for the test galaxy and for the density of the ICM it is moving through. We refer to such tests, whose ram pressure is constant and unchanging as `wind-tunnel' tests. It should be noted that this situation is artificial for real galaxies whose infall into a cluster which are subjected to both changing wind velocities and densities (discussed further in Section \ref{clustermodel}).

We choose an arbitrary fixed ICM density of $5.0 \times 10^{-4}$ cm$^{-3}$, corresponding to an intermediate density in the Virgo cluster. For example, a galaxy would encounter such densities at $\sim 600$ kpc from the cluster centre, assuming the reasonable $\beta$-model used for cluster infall tests in this study (described in full in Section \ref{clustermodel}). We subject our standard galaxy model to an ICM wind of constant velocity $v$ with a face-on inclination. $v$ is varied from 300-4000 km s$^{-1}$ where complete stripping is predicted at 3500 km s$^{-1}$. For a complete list of the wind-tunnel tests used with the standard model, see the upper entry of Tab. \ref{windteststab}.

The model galaxy is subjected to the ICM wind for 0.75 Gyr. This duration is chosen as it allows sufficient time for gas that has been unbound to be accelerated away from the stellar disk, at which point the gas truncation radius is measured. The duration is also physically motivated. \cite{Trentham2002} give the crossing time of the Virgo cluster as one tenth of a Hubble time. Therefore 0.75 Gyr represents a rough time-scale for which a Virgo cluster galaxy might experience significant ram pressures.

A comparison between the analytical predicted $R_{\rm{trc}}$ of Equation \ref{truncrad}, and the measured gas truncation radius is shown in Fig. \ref{gandgcomp}. The hatched area marks the region in which gravity is not fully resolved in the model. In general the model matches the predicted truncation radius reasonably well, although there may be some over-stripping as we approach the resolution limit.

We see little indication of the under-stripping that might be expected if the dark matter halo of our model was substantially contributing to the restoring gravitational force of the galaxy acting on the gas disk against ram pressure. This supports our use of Eqn. \ref{truncrad} to predict the gas truncation radius.

\subsection{A cluster model}
\label{clustermodel}
From Eq. \ref{RPSpress}, the instantaneous ram pressure that a gas cloud feels is a function of both the velocity of the cloud, and the density of the ICM it moves through at that instant. In Section \ref{results}, we refer to wind-tunnel tests. In these highly idealised tests, the ICM density $\rho_{\rm{ICM}}$, and ICM wind velocity $v$ is fixed, producing a constant and unevolving ram pressure on a galaxy model's disk of magnitude $\rho_{\rm{ICM}} v^2$.

Instead, a galaxy that orbits in a real galaxy cluster faces strong time-evolution of the ram pressure. The density of the ICM rises significantly with decreasing radius within the cluster. Additionally, the gravitational potential well of the cluster causes a galaxy to have highest orbital velocity through the densest ICM, when at the pericentre of the orbit. Therefore ram pressure is expected to be strongly peaked close to pericentre. Hence the evolution of the ram pressure depends on the shape of the ICM density profile, and the potential well of the cluster. Under the assumption that the ICM gas is isothermal and in hydrostatic equilibrium within the potential well of the cluster, and the gas has negligible contribution to the cluster potential well, the following equation holds:

\begin{figure}
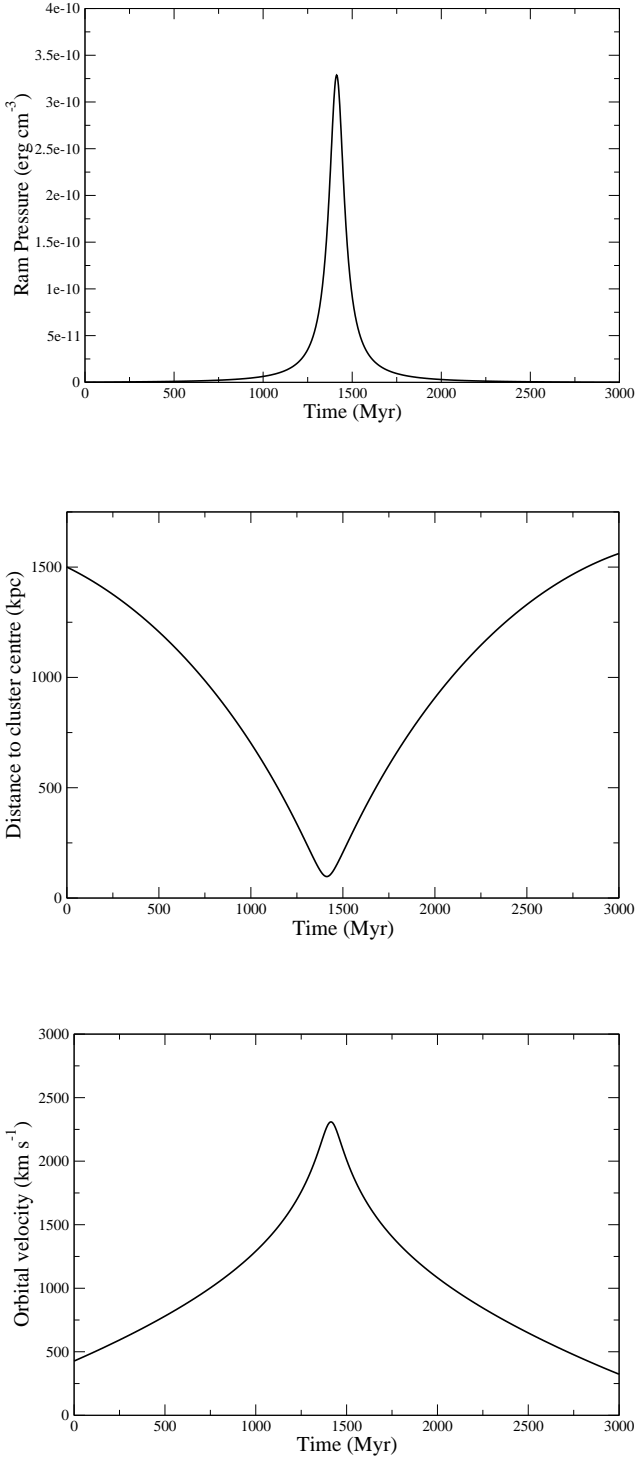

\begin{center}$
\begin{array}{c}  
\includegraphics[width=3.3in]{pressevol.eps} \\ 
\\
\\
\\
\includegraphics[width=3.3in]{radiusevol.eps} \\
\\
\\
\\
\includegraphics[width=3.3in]{velevol.eps} \\
\end{array}$
\end{center}
\caption{Time evolution of key parameters along the orbit. (Upper panel) Ram pressure, (Centre panel) Radius within the cluster, and (Lower panel) orbital velocity}
\label{orbitquantities}
\end{figure}

\begin{equation}
a(r)=-\frac{k_B T}{\mu m_p} \frac{d}{dr} \bigg({\rm{ln}} (\rho_{\rm{ICM}})\bigg)
\end{equation}

where $a(r)$ is the acceleration due to the potential well at radius $r$, $k_B$ is Boltzmann's constant, $T$ is the isothermal gas temperature, $\mu$ is the mean molecular mass (assumed equal to 0.6), $m_p$ is the mass of a proton, and $\rho_{\rm{ICM}}$ is the density profile of the ICM. For $\rho_{\rm{ICM}}$, we assume a $\beta$-profile:

\begin{equation}
\label{rhoICM}
\rho_{\rm{ICM}}(r) = \rho_0 \left(1+\frac{r^2}{r_{\rm{ICM}}^2}\right)^{(-3/2)\beta}
\end{equation}

Following the cluster model C1 in \cite{Roediger2007}, we assume $T=4.7\times 10^{7}$K, $\rho_0=2\times 10^{-26}$g cm$^{-3}$, $\beta=0.5$, and $r_{\rm{ICM}}$ is 50 kpc. As noted in the same Roediger \& Bruggen paper, this is similar to the Virgo cluster (\citealp{Matsumoto2000}), but less centrally concentrated. This fully defines the cluster model's gravitational potential well, and the radial dependence of ICM densities.

\section{Results}
\label{results}
\subsection{A galaxy infall within the cluster}
\label{clusterinfall}

As previously discussed, the ram pressure experienced by a galaxy infalling into a cluster is a strong function of time. We model the changing ram pressure in the same manner as \cite{Vollmer2001}.

The ram pressure wind is assumed to be mono-directional throughout the galaxy's orbit in the cluster. This is idealised, but for the plunging orbits we consider, the galaxy trajectory near the cluster centre are quasi-linear, and for now we wish to restrict our study to purely face-on stripping (i.e. the wind is inclined orthogonal to the plane of the disk).

Using a single particle integrator, we calculate orbits within the cluster potential well. Initially the particle is positioned at 1.5 Mpc from the cluster centre, at an initial radius where the ICM density is very low resulting in initially negligible ram pressure. We arbitarily choose the particle's initial velocity vector such that it has a 400 km s$^{-1}$ radially inwards component and a 150 km s$^{-1}$ tangential component. This results in a plunging orbit. The evolution of ram pressure, radius within the cluster, and orbital velocity that would result from this orbit can be seen in Fig. \ref{orbitquantities}.

The galaxy model is then located at the origin in simulation space, and is subjected to a time-evolving ram pressure matching that of the upper panel in Fig. \ref{orbitquantities}. We note that this technique of modelling the effects of the time-evolving ram pressure that a galaxy might feel on an infall through a cluster is identical to that of \cite{Vollmer2001}.

For this test we deliberately choose a galaxy model that matches the standard model in all parameters, except we choose a gas-to-stellar mass ratio of 0.1. It is therefore considerably less gas-rich than the standard model, with a gas fraction matching massive late-type galaxies (\citealp{Gavazzi2008}).

\begin{figure*}
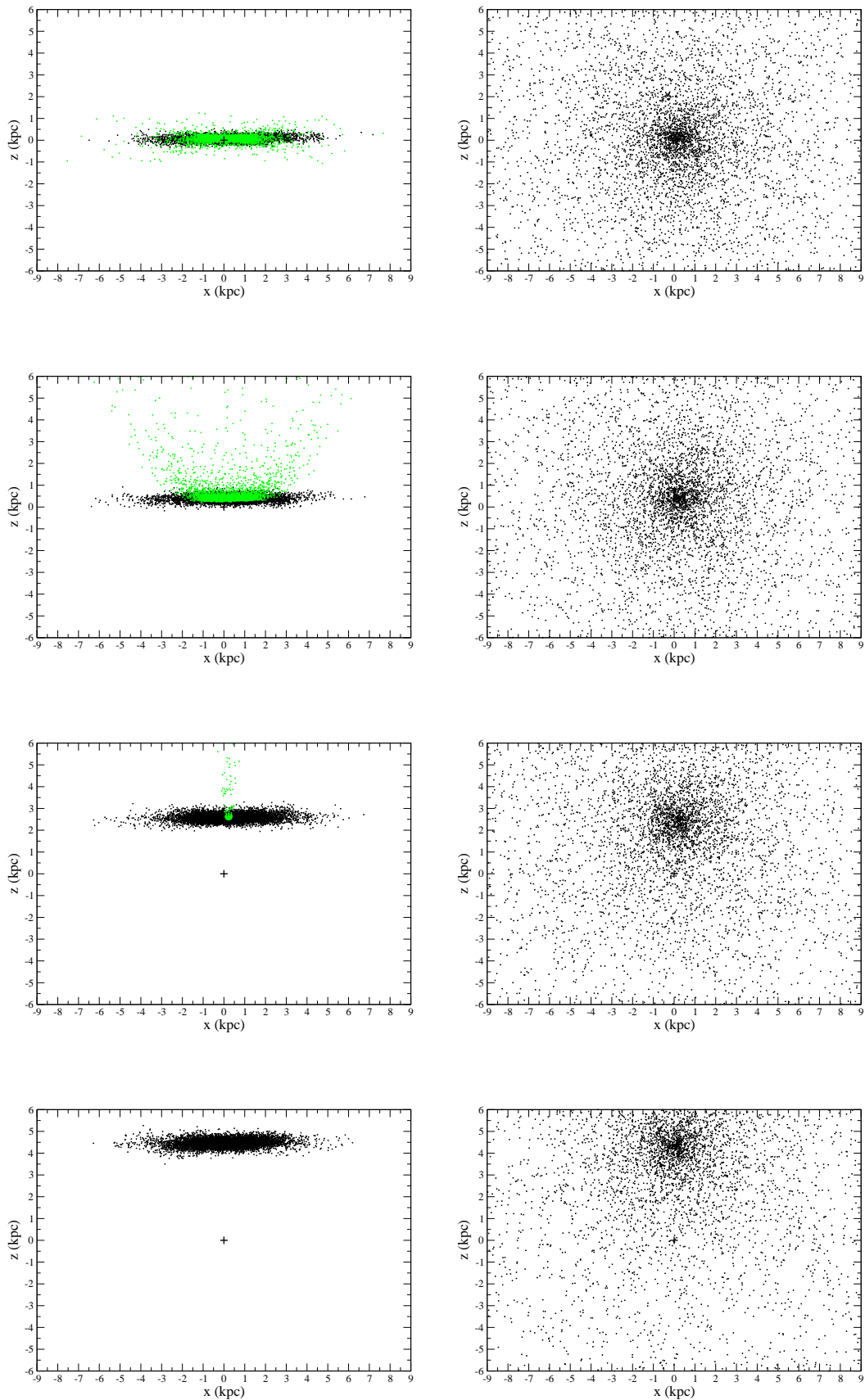

\begin{center}$
\begin{array}{ccc}  
\includegraphics[width=2.6in]{gasnstars_1.eps} & &
\includegraphics[width=2.6in]{dm_1.eps} \\ 
\\
\\
\\
\includegraphics[width=2.6in]{gasnstars_20.eps} & &
\includegraphics[width=2.6in]{dm_20.eps} \\
\\
\\
\\
\includegraphics[width=2.6in]{gasnstars_50.eps} & &
\includegraphics[width=2.6in]{dm_50.eps}\\
\\
\\
\\
\includegraphics[width=2.6in]{gasnstars_80.eps} & &
\includegraphics[width=2.6in]{dm_80.eps}\\
\end{array}$
\end{center}
\caption{Snapshots of galaxy components, evolving in time from top row to bottom row. Top row is 0.00 Myr, second row is 500 Myr, third row is 1400 Myr, and bottom row is 3000 Myr. Left column shows star particles (dark,black), and gas particles (light,green). Right column shows dark matter particles (dark,black). A plus symbol marks the original centre of the galaxy halo and disk at 0.00 Myr.}
\label{snapshots}
\end{figure*}

In Fig. \ref{snapshots} we plot x-z projections of the evolution of the stellar and gas component; dark (black) and light (green) particles respectively in left panel), and the dark matter (dark (black) particles in right panel). The ram pressure wind blows towards positive z in these figures. Initially ram pressure is weak when the galaxy is at low velocities and in the low density ICM regime. After 0.5 Gyr (the second row), the gas disk is mildly truncated to $\sim 2-3$ kpc. Stripped gas is blown in the direction of the wind creating a conical tail of stripped gas.
At 1.4 Gyr (the third row), the model is near to orbit pericentre and is moving rapidly through a dense ICM. The gas disk is heavily truncated, with gas remaining in the inner $\sim 0.3$ kpc only. After 3.0 Gyr the model has effectively passed orbit pericentre, is entirely stripped of its gas, and has returned to the outskirts of the cluster.

The key point to note is throughout the infall into the cluster the stellar disk and remaining gas disk of the galaxy is dragged away from it's original location in the direction of the wind (marked with a cross symbol in the panels of Fig. \ref{snapshots}). At 3.0 Gyr, the stellar disk has been shifted $\sim 4.5$ kpc. Comparing with the right column, it can be seen that the central dark matter surrounding the stellar disk is also ram pressure dragged. The central cusp of the dark matter halo remains superimposed on the location of the stellar disk and remaining gas disk, but is also offset from it's original location by $\sim 4.5$ kpc.

\begin{figure}
  \centering \epsfxsize=8.0cm \epsfysize=8.0cm \epsffile{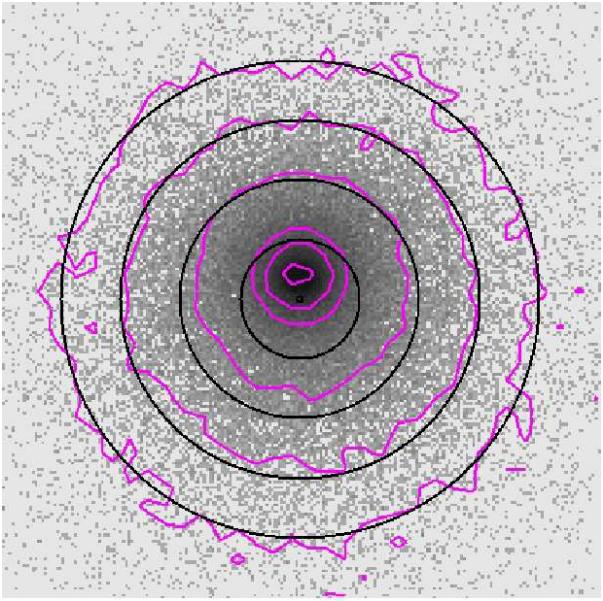}
  \caption{Grey-scale surface-density plot of the dark matter halo at 3.0 Gyr. The grey-scale image is overlayed with surface-density contours (light/magenta curves) at 50, 10, 5, 1, 0.3, and 0.1 M$_\odot$ pc$^{-2}$ (from inner contour to outer respectively). Centred on the original location of the halo before ram pressure stripping, are circles (black) with radii of 0.6, 12, 24, 36 and 48 kpc (from the inner to outer circle respectively). The ram pressure wind blows from the bottom to the top of the plot. Central dark matter is the most significantly dragged, and the strength of the deformation decreases with radius. The outer halo remains largely unaffected. This figure is 120 kpc on each side. A colour version is available online.}
\label{dmconts}
\end{figure}

Dark matter in the outer halo appears largely unaffected and remains spherical, and centred on its original location. However the inner regions of the halo have been dragged off-centre with respect to the outer halo. In Fig. \ref{dmconts} we show a grey-scale surface-density plot of the dark matter halo at 3.0 Gyr (corresponding to snapshot in the bottom panel of Fig. \ref{snapshots}). We overlay the grey-scale image with surface-density contours (light/magenta curves) at 50, 10, 5, 1, 0.3, and 0.1 M$_\odot$ pc$^{-2}$. Centred on the original location of the halo, we additionally mark circles (black) with radii of 0.6, 12, 24, 36 and 48 kpc. This demonstrates the extent to which the originally spherical halo has been deformed. The outer halo ($r>36$ kpc) appears largely unaffected. At intermediate radius ($r\sim$24 kpc) a halo deformation, in the direction of the ram pressure wind (towards the top of the figure), becomes apparent. The extent of the deformation increases with decreasing radius, and is most apparent in the inner 12 kpc of the halo.

Along with the ram pressure dragging of the stellar disk and dark matter, there is the tentative hint that the stellar disk is not entirely unaffected during it's infall through the cluster. Comparing the final stellar disk to the initial stellar desk in Figure \ref{snapshots}, there is a hint of mild disk thickening, although in this case it is not significant. This indicates that dark matter potential well surrounding the disk has not been heavily altered, as the disk remains largely unaffected. In the following section we concentrate on explaining and understanding the dragging of the stellar disk and surrounding dark matter by ram pressure. We return to the issue of stellar disk thickening in a later section.

\subsection{A simple analytical description}
We are not the first to note that ram pressure can result in dragging of the stellar disk of a galaxy undergoing ram pressure. \cite{Schulz2001} report that the stellar disk of their galaxy model is dragged by 2 kpc. Considering an annular region in the outer part of the disk, they suggest the gas disk in this annulus is analogous to a solar sail, which is dragging a payload - the stellar disk. We offer an alternative but complimentary approach to an analytical understanding of ram pressure drag. As we will demonstrate, our approach provides us with the drag force directly, and predicts it's dependency on ram pressure strength and disk properties.

First let us assume a galaxy model containing a dark matter halo, and exponential disk of gas and stars. The gas disk has mass $m_{\rm{g}}$ and scalelength $R_{\rm{g}}$. The stellar disk has mass $m_\star$ and scalelength $R_\star$. We now assume that the galaxy is in motion through an ICM of density $\rho_{\rm{ICM}}$, at a constant velocity $v$.

In this scenario, according to the Gunn and Gott criteria the gas disk will be truncated to the truncation radius $R_{\rm{trc}}$ given by Eqn. \ref{truncrad}. At this radius the self-restoring gravity of the galaxy is sufficient to maintain the gas against the ram pressure.

The remaining gas disk is subjected to a continuous force that is simply the product of the pressure of the wind with the area of the disk: $F_{\rm{drag}}=\rho_{\rm{ICM}} v^2$A where A is the area of the disk. For a truncation radius $R_{\rm{trc}}$, A$=\pi R_{\rm{trc}}^2$. Substituting in Eq. \ref{truncrad} we have, 

\begin{equation}
\label{fdrag}
{\rm{F}}_{\rm{drag}}=\rho_{\rm{ICM}} v^2 \pi\bigg[\left(\frac{1}{R_\star}+\frac{1}{R_{\rm{g}}}\right)^{-1} {\rm{ln}}\left(\frac{2 \pi \rho_{\rm{ICM}}v^2 R_{\rm{g}}^2 R_{\rm{\star}}^2}{G m_{\rm{\star}} m_{\rm{g}}}\right)\bigg]^2{\rm{.}}
\end{equation}
\noindent

To stop the remaining gas disk accelerating away, the galaxy must have sufficient self-gravity to overcome the ram pressure. The central halo, stellar disk and remaining gas disk of the galaxy provide sufficient gravitational force to accomplish this. However from Newton's third law, it is inevitable that in providing this restoring force, these components of the galaxy must also feel an equal and opposite force in the direction of the wind.

In other words, the drag force on the truncated gas disk does not disappear, simply because the gas disk is held in place by the self-gravity of the galaxy. Instead it is passed to all components of the galaxy through their mutual gravity. The communication of the drag force to the stellar disk and dark matter of a galaxy is also discussed in \cite{Fujita2006} - if the gas disk is dragged, both the dark matter and stellar disk will pull back.

The ram pressure drag effect could not have been seen in numerical models of ram pressure that assume a fixed, static analytical potential for the dark matter and stellar disk component of their model disk galaxies. Ram pressure drag of dark matter and stellar disks occurs because in a live dark matter, the halo must respond to the applicaton of additional forces. Given sufficient time, the continuous drag force of ram pressure can cause sufficient acceleration of the central dark matter to pull it off centre as demonstrated in Fig. \ref{dmconts}.

Eq. \ref{fdrag} provides a simple quantification of the size of 
${\rm{F}}_{\rm{drag}}$ for an exponential disk of gas and stars. However the resulting response of the central dark matter to this forcing is difficult to predict using only simple analytical considerations. The degree of acceleration of the central cusp will likely depend on the mass of dark matter that is effected by ram pressure drag. Furthermore, an offcentre cusp within an unaffected outer halo must feel additional restoring forces from the outer halo. It's resulting motion to the ram pressure drag force is unlikely to have a simple analytical solution. For this reason, we rely on numerical simulations to quantify the effect of ram pressure drag on the stellar disk and central dark matter halo displacement.

However, using the numerical simulations, we will demonstrate how Eq. \ref{fdrag} appears to successfully predict, to first order, the drag force and its dependency on ram pressure strength and disk properties. 

\subsection{Wind-tunnel tests on the standard model}

We subject the standard model galaxy to wind-tunnel tests in the same manner as described in Section \ref{gandgtests}, and assuming the same moderate ICM density (5.0$\times$10$^{-4}$ cm$^{-3}$). 

The dark matter, stellar disk and remaining gas disk are seen to always remain superimposed in our models (for example, see Fig. \ref{snapshots}). Therefore by measuring the location of the centre-of-mass of the stellar disk, we simultaneously measure the displacement by which the central halo dark matter and the stellar disk have been dragged away from their initial location at the centre of the dark matter halo. 

Herein, we shall refer to the size of this displacement as the {\it{`cusp displacement'}}, although it should be remembered that it is also the displacement of the stellar and remaining gas disk away from their original position. 

Initially the outer halo dark matter is unaffected by cusp dragging and remains spherically symmetric. Therefore cusp displacement can be regarded as the distance that the central dark matter cusp (and disk) moves off-centre with respect to the outer halo.

\begin{figure}
  \centering \epsfxsize=8.0cm \epsfysize=6.5cm \epsffile{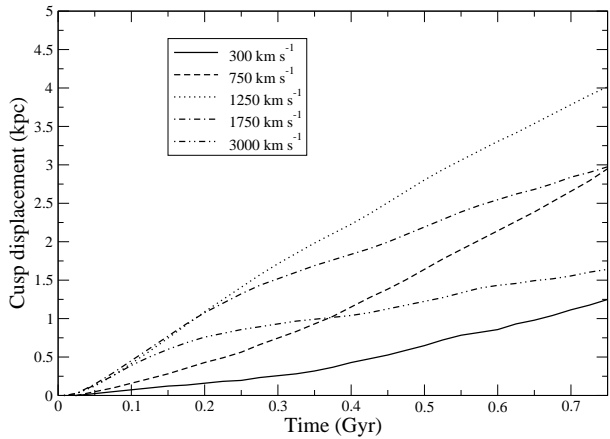}
  \caption{Time-evolution of the {\it{`cusp displacement'}} - the position of the cusp, stellar disk and remaining gas disk with respect to their original position. Wind-tunnel ram pressure simulations are used with a range of wind-speeds (as indicated in the key). At low and high ram pressures, the cusp and disk is less significantly dragged. There exists a range of ram pressures, at intermediate wind speeds, when cusp and disk dragging occurs the most.}
\label{varywind}
\end{figure}

In Fig. \ref{varywind}, we plot the time evolution of cusp displacement. Each curve represents a specified ICM wind velocity. At 300 km s$^{-1}$, the cusp displacement slowly but steadily grows with time. After 0.75 Gyr, the total cusp displacement is 1.25 kpc. At 750 km s$^{-1}$, cusp displacement occurs more rapidly, with a larger final cusp displacement. Peak displacement occurs between wind velocities $\sim$1000-1500 km s$^{-1}$. At $v$=1250~km~s$^{-1}$ total cusp displacement is $\sim$4.0 kpc after 0.75 Gyr. As the wind velocity is increased further, the final cusp displacement does not increase further, and begins to drop. 

As we will show, our analytical expression for the drag force shown in Eq. \ref{fdrag} qualitatively predicts this behaviour. For clarity, we repeat that $F_{\rm{drag}}=\rho_{\rm{ICM}} v^2$A where A is the area of the disk. At sufficiently high ram pressures, it is therefore inevitable that the area of the gas disk will become sufficiently small that the drag force upon it must decrease . As an extreme case, a galaxy undergoing very high ram pressures that completely strip the gas, will no longer present any surface area to the wind. At this point $F_{\rm{drag}} \rightarrow 0$. It is therefore predicted that the drag force on this disk will peak at some intermediate value of ram pressure; between a low pressure that allows a large gas disk area but is too weak to result in strong dragging, and a high ram pressure that reduces the gas disk area so much, that there is in sufficent area on which the ram pressure can act.

Using Eq. \ref{fdrag}, for the standard model we can calculate the value of $F_{\rm{drag}}$ for varying ram pressures (for the fixed ICM density of the wind-tunnel tests, simply varying the ICM wind speed, is equivalent to varying the ram pressure). This is plotted as the solid line in Fig. \ref{dragforce}.

A peak in the drag force is predicted for wind velocities between $\sim$1000-1500 km s$^{-1}$. At wind velocities smaller or larger than this value, the drag force is expected to decrease. As mentioned in the previous section, a knowledge of the drag force on central dark matter does not necessarily directly translate into predictions of how the cusp will move under this forcing. However the cusp displacement indeed peaks near 1000 km s$^{-1}$ where the drag force is predicted to peak. At velocities greater than 1750 km s$^{-1}$ or lower than 750 km s$^{-1}$, the cusp displacement lowers, in the same qualitative manner as the drag force is predicted to lower.

It is interesting to note the initial motion of the cusp at wind speeds significantly beyond the predicted peak drag force. For example, see the $v=3000$ km s$^{-1}$ curve on Fig. \ref{varywind}. Initially cusp motion is fast at high ram pressures, then tapers off. This behaviour is indicative of the finite time required for stripped gas to be accelerated away from the stellar disk in the galaxy models (see \cite{Roediger2007} for a more complete discussion of this phonomena). The standard Gunn and Gott stripping criteria assumes instantaneous stripping of any gas located at disk annuli where the self-gravity of the disk weaker than the ram pressure. However, in reality this process is not instantaneous, and during this period, the stellar disk and dark matter halo is subjected to heightened drag forces until the gas disk has been stripped to it's final truncation radius. As our formulation of F$_{\rm{drag}}$ in Eq. \ref{fdrag} is derived from the instantaneous Gunn and Gott Equation, it does not account for this initial burst of strong dragging, that appears most significant at high ram pressures.

\subsection{Dependency of dragging on disk galaxy properties}
\label{diskgalpropssection}
\subsubsection{Analytical predictions}
So far we have only considered the effects of ram pressure drag on a single fixed galaxy model. Now we consider alternative disk galaxy models.
In all the following disk galaxy models, we restrict our study to that of a 10$^{10}$M$_\odot$ dark matter halo of equal mass to that of the standard model.

We first consider a `heavy disk' model. For this model, the disk matches that of the standard model except it's mass is doubled. Therefore in this model, the disk mass is 10$\%$ of the dark matter halo mass.

We also consider a `small disk' model. In this case, the disk matches the standard model except it's exponential scalelength is reduced from 0.73 kpc to 0.56 kpc. In the \cite{Mo1998}, this is predicted if the spin parameter $\lambda$ is chosen to be 0.04. In cosmological simulations, $\lambda$ is found to have a log-normal distribution that peaks near to 0.05 and has a range varying from 0.02 to 0.08. Therefore our choice of $\lambda$ for the 'small disk' model is entirely reasonable.

We also consider two disk models that are identical to the standard model except we vary the gas-to-stellar mass of their disks. Both are significantly less gas-rich than the standard model which contains equal gas to stellar mass. The `g/s=0.3' model has a disk that is 76$\%$ stars and 24$\%$ gas. The `g/s=0.1' model has a disk that is 90$\%$ stars and 10$\%$ gas. Note that `g/s' refers to the mass ratio of gas to stars in their disk. The `g/s=0.1' model is identical to that used in the cluster infall simulation described in Section \ref{clustermodel}.

Finally, we consider a `low concentration' model, in which the model is identical to that of the standard model, except the dark matter halo is significantly less concentrated with $c=5$. The ranges of halo concentrations found in cosmological dark matter simulations are found to follow a scattered trend towards higher concentrations in lower mass halos (\citealp{Lokas2001}). A concentration of $c=5$ is reasonable within the significant scatter observed in this trend.

\begin{figure}
  \centering \epsfxsize=8.5cm \epsfysize=6.0cm \epsffile{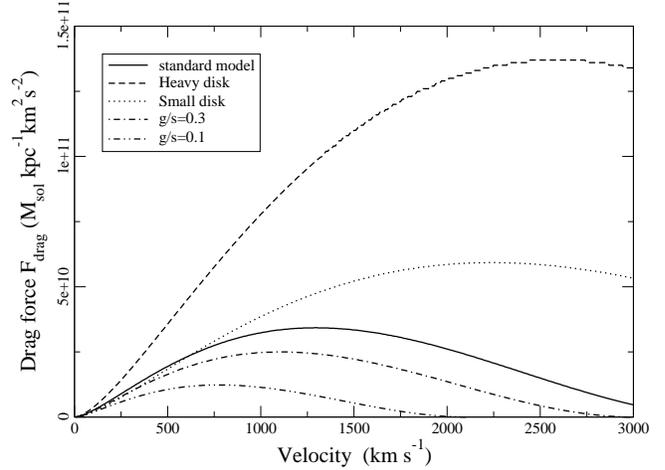}
  \caption{The analytical formulation of ram pressure drag is used to predict the drag force at a range of wind speeds for the standard model, heavy disk, small disk, and reduced gas fraction models (different models are indicated by the curves in the key). In comparison to the standard model; Lowering the gas fraction reduces the drag force at all velocities. The smaller disk model is predicted to have similar drag forces at low wind speeds, but increasingly stronger drag forces at higher wind speeds. The heavy disk model has the strongest ram pressure drag at all wind speeds. The standard model curve is identical for the low halo concentration model as the disk is identical.}
\label{dragforce}
\end{figure}

The analytical prediction for the magnitude of the drag force in each of these galaxy models, including the standard model is shown on Fig. \ref{dragforce}, as indicated in the key. It should be noted that the standard model and the low concentration model share the same curve, as their disk properties are identical. The heavy disk model is predicted to feel the strongest drag force for the whole range of wind velocities considered. Additionally, it's peak drag force occurs at much higher wind velocity than the standard model. The small disk model has a higher surface density than the standard model. At high wind velocities, this results in a prediction of stronger ram pressure drag force than the standard model. However at wind velocities $<1000$ km s$^{-1}$, their drag forces should be comparable. As might be intuitively expected, lowering the gas fraction of this disk should result in weaker drag forces - the mass that tows the stellar disk and halo cusp has been reduced. The difference between predicted drag forces of the standard model and the low gas fraction models becomes more prominent at higher wind velocities. The lower the gas fraction, the more easily the disk is stripped to small truncation radii (or altogether). This results in a peak drag force that occurs at lower wind velocities with decreasing gas fraction.

\begin{figure}
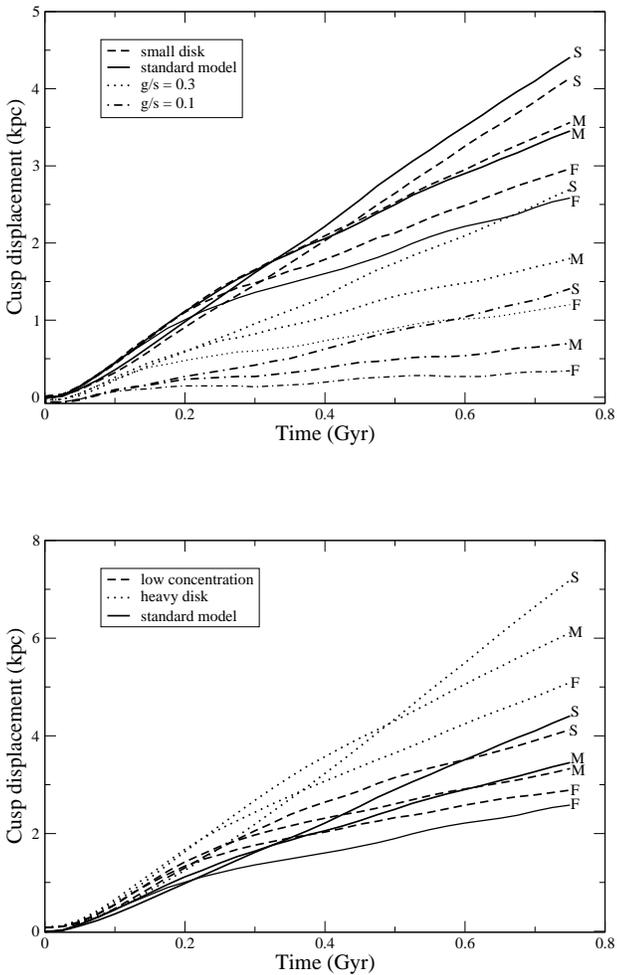

\begin{center}$
\begin{array}{c}  
\includegraphics[width=3.2in]{diskpropdrag2.eps} \\ 
\\
\\
\\
\includegraphics[width=3.2in]{diskpropdrag1.eps} \\
\end{array}$
\end{center}
\caption{Time-evolution of cusp displacement in wind-tunnel tests for a 2000 km s$^{-1}$ fast wind (F), a 1500 km s$^{-1}$ medium wind (M), and a 1000 km s$^{-1}$ slow wind (S). Wind-speed labels are indicated at the end of each curve. Line-styles indicate different disk models as shown in the key. Results for the small disk and reduced gas-fraction models are in the upper panel. Results for the low concentration and heavy disk model are in the lower panel. For ease of comparison, we put standard model is shown in both plots.}
\label{actualdiskpropdrag}
\end{figure}

\subsubsection{Wind tunnel tests}
Having discussed the analytical predictions for drag forces for each disk model, we now test the actual response of each model to ram pressure drag using the toy model of ram pressure. We use wind tunnel tests matching those used in the previous section, but this time applied to our new disk models. Once more, we measure the cusp displacement in each wind tunnel test over a period of 0.75 Gyr. We remind the reader that the cusp displacement is the distance at which the dark matter halo cusp, stellar disk and remaining gas disk are dragged away from their original location. 

For these tests we subject each galaxy to different wind speeds; slow (1000 km s$^{-1}$), medium (1500 km s$^{-1}$), and fast (2000 km s$^{-1}$). The results are presented in Fig. \ref{actualdiskpropdrag}. We indicate the wind speed with a `F' (fast),`M' (medium), or `S' (slow) at the end of each curve on this figure. For reference, a summary of all the wind tunnel tests conducted in this paper is provided in Tab. \ref{windteststab}.

\begin{table}
\centering
\begin{tabular}{|c|c|}
Galaxy model & Wind speed (km s$^{-1}$) \\\hline\hline
\multirow{4}[4]{*}{Standard} 
&300,500,750,1000 \\
&1250,1500,1750 \\
&2000,2250,2500 \\
&3000,3500,4000 \\ \hline
Small & 1000,1500,2000 \\ \hline
Heavy & 1000,1500,2000 \\ \hline
Low concentration & 1000,1500,2000 \\ \hline
Gas-to-star ratio=0.1 & 1000,1500,2000 \\ \hline
Gas-to-star ratio=0.3 & 1000,1500,2000 \\
\end{tabular}
\caption{A complete list of all wind-tunnel tests conducted in this study, and the model galaxies they were conducted on}
\label{windteststab}
\end{table}

The upper panel of Fig. \ref{actualdiskpropdrag} shows the standard model versus the small disk, and reduced gas fraction models. It is analytically predicted (see Fig. \ref{dragforce}) that the small disk will suffer similar drag forces as the standard model at the slow wind velocity. However as the wind velocity is increased beyond this, the small disk will suffer increasing larger drag forces than the standard model. To first order, the magnitude of cusp displacement appears to follow strength of the predicted drag force. There is similar cusp displacement at low and medium wind velocities, but larger cusp displacement for the strong wind case. Comparing the low gas fraction models with the standard model, we see reduced cusp motion with reducing gas fraction. This is also expected as reduced gas fraction results in reduced disk drag at all velocities.

The lower panel of Fig. \ref{actualdiskpropdrag} shows the standard model versus the heavy disk and low concentration model. We first discuss the comparison between the standard model and the heavy disk. It is analytically predicted (see Fig. \ref{dragforce}) that the heavy disk suffers significantly stronger drag forces at all wind velocities. From the simulation, we see this results in the heavy disk suffering the most cusp displacement at all wind velocities. Once more, a stronger predicted drag force results in a relatively greater cusp displacement.

In the low concentration halo test, the reduced concentration of the halo reduces the mass of dark matter surrounding the disk. This effectively reduces the analogous payload that the gas disk must drag. This results in an initial motion of the cusp that moves more rapidly than the standard model. For the fast wind, this results in a final cusp displacement that is greater than the standard model. 

\begin{figure}
  \centering \epsfxsize=8.0cm \epsfysize=6.0cm \epsffile{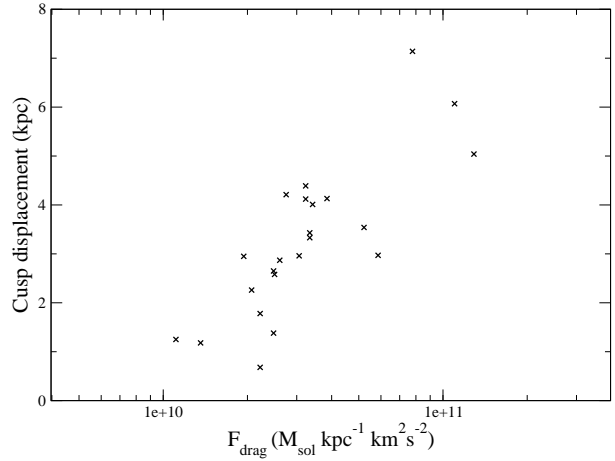}
  \caption{Cusp displacement of wind tunnel tests versus the analytically predicted drag force. With increasing drag force, we see increasing cusp displacement. However the proportionality is not linear (note logarithmic x-axis scale). There is also scatter in the trend - especially at the highest drag forces.}
\label{fdrag_displ}
\end{figure}

\subsubsection{Limitations of the analytical treatment}
However, when the gas disk is not truncated in the manner predicted by the Gunn $\&$ Gott formula, an increase in predicted drag force does not always result in an increase in the actual cusp displacement meaured in the model.

For example, we note that the heavy disk is predicted to suffer increasingly strong drag forces, in comparison to the standard model, up to 2000 km s$^{-1}$ and beyond. In the heavy disk simulation the fastest wind in fact causes the least cusp displacement. Also in the low concentration test for the medium and slow wind, the increased dragging due to reduced dark-matter mass about the disk is short-lived, and the final cusp displacement is similar to that of the standard model.

In these cases, the true ram pressure drag force is likely to be less than the predicted analytical drag force. The ram pressure drag force can become very sensitive to small deviations from the predicted truncation radius $R_{\rm{trc}}$ when the real truncation radius is small. For example, the standard model has a predicted $R_{\rm{trc}}=0.41$ kpc for the strong wind. Fig. \ref{gandgcomp} demonstrates that our simple ram pressure model can result in a real truncation radius $\sim$0.1 kpc smaller. As F$_{\rm{drag}}$ is proportional to the area of the truncated disk A$=\pi R_{\rm{trc}}^2$, a reduction of only 0.1 kpc in the truncation of the gas disk results in a drag force that is almost halved.

Furthermore, the real response of the model's gas disk to strong forcing may not be as idealised an in the Gunn $\&$ Gott formulation. \cite{Schulz2001} note that while the gas disk is dragging the stellar disk (and central dark matter) it effectively feels an increased surface density. In the simulations, we see that this can result in enhanced disk barring. Bar enhancement occurs for all the disk models but is especially strong for the heavy disk and low concentration model. The bars are dense, and reduce the cross-sectional area of the disk to the wind. Therefore the analytically predicted drag force may be too small. Also the disk cross-sectional area can become far from circular in shape once the outer gas is stripped to the perimeter of the bar, whereas a circular cross-section to the wind is intrinsically assumed in Eq. \ref{fdrag}.

\begin{figure*}
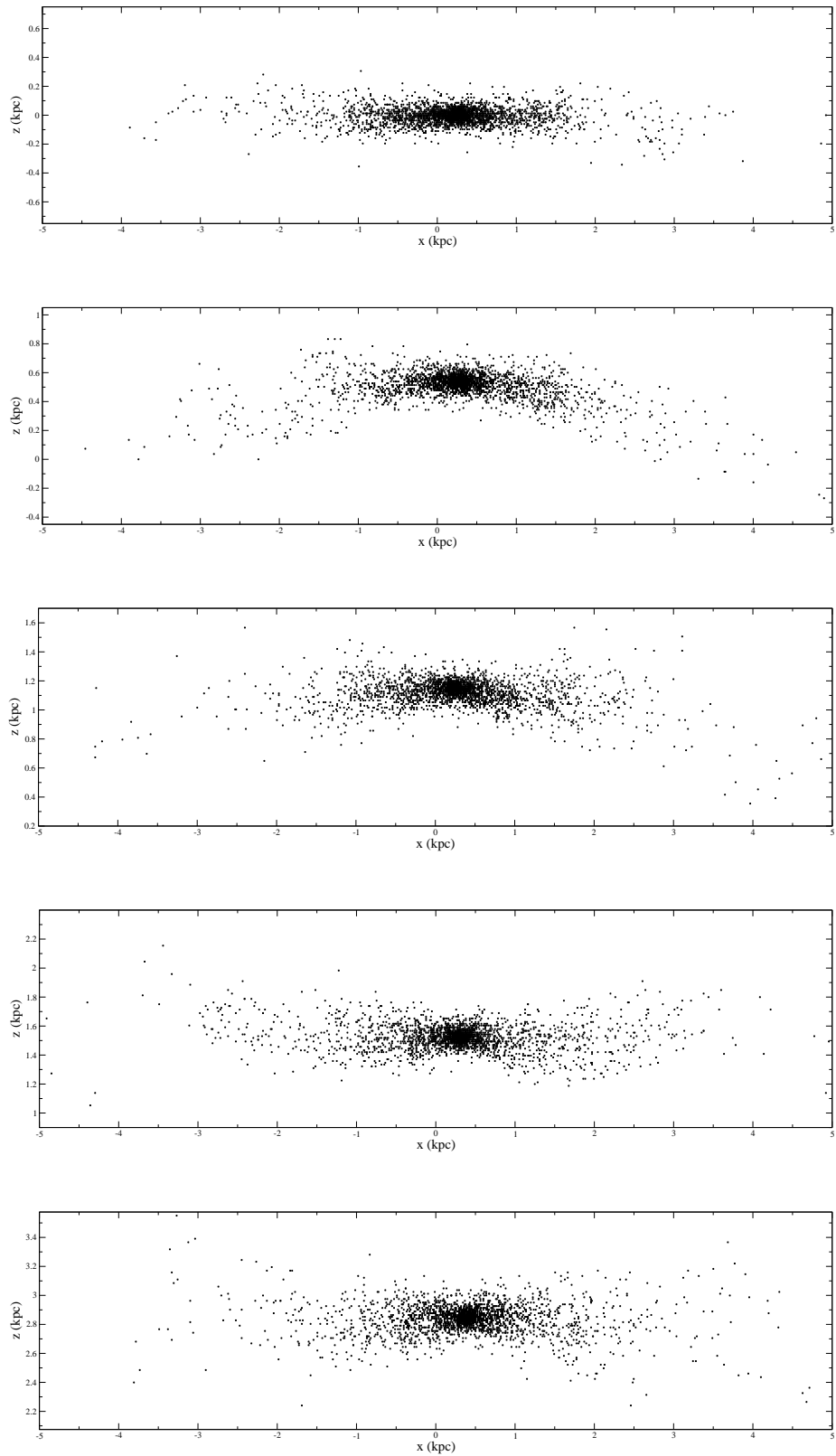

\begin{center}$
\begin{array}{c}  
\includegraphics[width=4.8in]{cone1.eps} \\ 
\\
\\
\includegraphics[width=4.8in]{cone2.eps} \\ 
\\
\\
\includegraphics[width=4.8in]{cone3.eps} \\ 
\\
\\
\includegraphics[width=4.8in]{cone4.eps} \\ 
\\
\\
\includegraphics[width=4.8in]{cone8.eps} \\ 
\end{array}$
\end{center}
\caption{Snapshots of the dynamical influence of a ram pressure wind tunnel test ($v$=1750 km s$^{-1}$) on the stellar disk of the standard model. For clarity, we do not include the gas particles in these panels, although in the simulation, gas is blown in a positive z-direction. Additionally, we present only a 0.5 kpc wide cross section of the stellar disk to emphasise the disk shape. The length of the axes is equal in all snapshot. From top to bottom, instant of snapshot is 0, 100, 200, 300 and 700 Myr of the wind-tunnel conditions. At 100 Myr, a conical distortion can be seen. At 200 Myr, the disk has almost returned to a flat shape, although at 300 Myr it has slightly over-shot being flat, and a milder inverted cone is visible. By 700 Myr, the disk is fully restabilised. Comparing the 0 Myr and 700 Myr snap shot, the final disk is clearly more thickened.}
\label{conemaker}
\end{figure*}

The extent to which barring occurs is expected to be dependent on the stability of a given model to Toomre instability. For example, the heavy disk model contains a gas disk that is closer to being Toomre unstable due to the high surface density of the disk. A massive dark matter halo provides stability to a disk, increasing it's Toomre stability parameter. However in the reduced concentration model, the disk is naturally closer to being Toomre unstable due to a reduced mass of dark matter surrounding the disk. Therefore the increased effective surface density that the gas disk experiences during ram pressure could result in more significant barring, and it is understandable that this is strongest for the heavy disk and low concetration model.

\cite{Schulz2001} use a more detailed treatment of the inter-stellar medium than our own, including a prescription for radiative cooling. They note that the increased effective surface density, that the gas disks of their models feel, results in the development of flocculent spiral structure. This spiral structure transfers angular momentum away from their disks, resulting in the gas disk being `annealed' to further ram pressure stripping. Additionally they comment that this behaviour requires a treatment of radiative cooling. We are not confident that we can describe the detailed gas dynamics and bar formation sufficiently realistically to draw firm conclusions using our simple isothermal model. We do not observe the development of floculent spiral structure, though \cite{Schulz2001} state that this behaviour requires a treatment of radiative cooling. However we note that the heavy barring observed in the heavy disk and small disk model will induce angular momentum loss that could result in a more compact disk.

In Figure \ref{fdrag_displ} we plot the cusp displacement at 0.75~Myr of the wind tunnel tests. We include points for all simulations where the truncation radius was fully resolved ($R_{\rm{trc}}>0.2$~kpc). We see the trend that might be expected - with increasing drag force there is increasing cusp displacement. 

However, the trend between dragging force and cusp displacement is not linear (note the logarithmic scale on the x-axis). As suspected, the resulting response of the central dark matter to this forcing cannot be calculated trivially. The degree of acceleration of the central cusp must depend on the mass of dark matter that is effected by ram pressure drag. Furthermore, an offcentre cusp within an unaffected outer halo must feel additional restoring forces from the outer halo. Therefore numerical simulations remain required to measure the cusp displacement of the galaxy due to ram pressure drag.

There is also some scatter in the trend. We suspect this is largely due to the response of the disk to the ram pressure that is not encapsulated in the Gunn and Gott Equation (Eq. \ref{GandGeqn}). This includes not instantaneous stripping (resulting in a brief burst of enhanced ram pressure drag). We note that this burst would not occur substantially for real galaxies who feel a more gradual build-up of ram pressures. This also includes barring of the gas disk - an effect that is most noteable in galaxy models experiencing the highest drag forces.

In summary, we conclude that our simple anaytical treatment is reasonably successful at predicting, {\it{to first order}}, the magnitude of the drag force, and it's dependencies on disk properties and ram pressure. However the tests additionally highlight that the analytical predictions will fail if the gas disk of a model is not truncated in the idealised manner described by the Gunn and Gott Equation (Eq. \ref{GandGeqn}).
\begin{figure*}
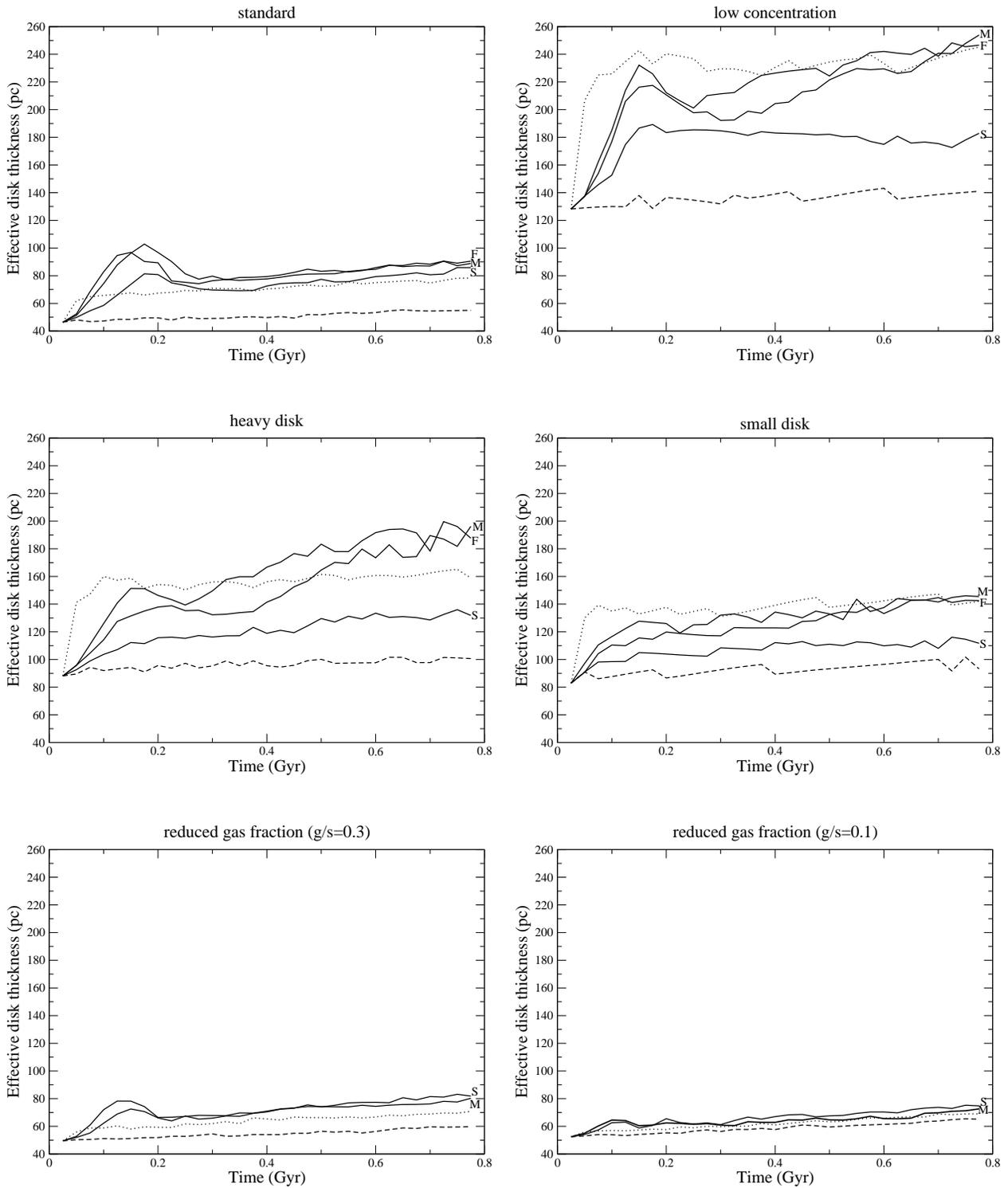

\begin{center}$
\begin{array}{cc}  
\includegraphics[width=3.2in]{zevol_standard.eps} &
\includegraphics[width=3.2in]{zevol_lowc.eps} \\ 
\\
\\
\includegraphics[width=3.2in]{zevol_heavy.eps} &
\includegraphics[width=3.2in]{zevol_small.eps} \\ 
\\
\\
\includegraphics[width=3.2in]{zevol_g2s0p3.eps} &
\includegraphics[width=3.2in]{zevol_g2s0p1.eps} \\ 
\end{array}$
\end{center}
\caption{Time-evolution of the effective disk thickness. Each panel is a different disk galaxy models (as indicated by the title of the panel). A dashed line represents the isolated/control model, and a dotted line represents the instantaneous gas removal model. Solid lines are wind-tunnel tests for a 2000 km s$^{-1}$ fast wind (F), a 1500 km s$^{-1}$ medium wind (M), and a 1000 km s$^{-1}$ slow wind (S). Wind-speed labels are indicated at the end of each solid-line. In most cases, disk thickening in wind-tunnel tests matches disk thickening in the instantaneous gas removal test, suggesting loss of the gas potential is the primary cause for disk thickening. Halo concentration is an additional parameter controlling the magnitude of disk thickening. The conical distortion can generally be seen as a temporary peak in effective disk thickness, typically lasting $<0.2$ Gyr.}
\label{zevol}
\end{figure*}

\subsection{Stellar disk heating}
So far, we have mainly considered the impact of ram pressure drag on the net displacement of the stellar disk and surrounding dark matter. In this section we shall consider the influence of ram pressure on the internal dynamics of the stars within the disk. The previous study by \cite{Schulz2001} states that the stellar disks of their models were largely unaffected internally, and were merely displaced. 

In our simulations, we see indications of disturbances to the stellar disk. Stellar disks can pass through a brief period of morphological reshaping, appearing mildly conical in shape (for example, see Fig. \ref{conemaker}). This reshaping occurs primarily as the dark matter does not respond to a centrally located drag force as a solid body. Dark matter located directly by the truncated gas disk moves first, with a small lag before more distant dark matter responds. We note that the distortion of the central dark matter must be fairly mild or the stellar disk would not continue to appear disk like. However, the centre of the stellar disk may be displaced before the outer stellar disk responds leaving a trailing outer disk, and resulting in the observed conical morphology. 

The conical distortion is most significant when the central dark matter is accelerated most strongly. In the wind-tunnel tests this occurs most dramatically, as effectively an initially isolated disk instantaneously meets the ICM wind. However, this situation is artificial as when a galaxy infalls into a cluster, the ram pressure builds up more gradually. To account for this, we repeat the cluster infall simulation described in Section \ref{clusterinfall} but for the standard model, low concentration, heavy disk, and small disk model, and for a variety of orbits. 

We do see a conical distortion in these simulations, as the ram pressure is very peaked for this orbit causing a sudden acceleration of the central dark matter. However, the overall effect is far milder than presented in Fig. \ref{conemaker}, due to a more gradual build up of ram pressure. Although it should be noted that the ICM may not be so smooth in real clusters as assumed in our cluster model. In particular, clusters that are still in the process of collapsing are very irregular (e.g. Virgo) and their ICM may be far from being isothermal, nor in hydrostatic equilibrium within the cluster potential well (\citealp{Tecce2010}).

The conical distortion is not long-lived in any of our simulations, lasting $<$200 Myr at most (see, for example, time-evolution in Fig. \ref{conemaker}). Following the conical reshaping, stellar disks are seen to return to a flattened shape, but with a thicker disk than prior to stripping. Disk thickening may occur as a result of the conical reshaping, and/or due to the removal of a significant fraction of the total disk's mass, once the gas component has been stripped. Here we analyse the relative contribution of the conical distortion and the loss of the gas potential to stellar disk thickening.

To quantify disk thickening, we measure the distance from the plane of the disk that contains half the total stellar mass. It is analogous to a galaxy effective radius, only measured vertically out of the plane of the disk. We refer to this quantity as the `effective disk thickness' herein.

We measure the evolution of the effective disk thickness in the wind tunnel tests described in Section \ref{diskgalpropssection}. Recall that disk galaxies of varying properties (standard, low concentration, heavy disk, small disk and gas fraction models) are subjected to a high, medium, and low ram pressure in these tests.

Thickening of the stellar disk occurs steadily with time, even in galaxy models that are evolved in isolation and are not ram pressure stripped. As noted in \cite{Moore1999}, there are a number of artificial sources of disk heating due to discreteness including finite time-stepping, artificially large dark matter particles, and softened gravity. However, using an isolated galaxy model as a control, evolved over the same time period, we can better understand the additional disk thickening that results from ram pressure stripping.

In order to quantify stellar disk heating that occurs purely from the reduction of the disk potential due to the removal of the gas component, we additionally model the extreme case of instantaneous removal of the gas. At a chosen instant, the gas component of our model vanishes instantaneously, and we meaure the resulting evolution of the effective disk thickness. These models are not subject to any stellar disk heating resulting from the conical distortion phase.

By comparison between evolution of the effective disk thickness in the control, the instantaneous gas removal, and the wind tunnel tests, we can quantify and separate the individual role of the loss of the gas potential, and the conical distortion to disk heating during ram pressure stripping. The results are shown in Fig. \ref{zevol}.

In each panel, solid lines represent the evolution of the effective disk thickness for the wind-tunnel tests. There are three solid lines for each model labelled to show the three wind speeds (`F' is fast, `M' is medium, `S' is slow) The dashed line is the isolated control galaxy model. The dotted line is instantaneous gas loss model. From left to right panel, starting at the top row and moving down, we show results for; the standard model, low concentration, heavy disk, small disk, and reduced gas fraction models (g/s=0.3 and g/s=0.1) respectively.

In each panel, the effective disk thickness can be seen to jump upwards from the isolated control model (dashed line), while undergoing ram pressure stripping in the wind-tunnel tests (solid lines). The size of the increase varies considerably between the models (different panels). 

In the wind tunnel tests (solid lines), disk heating could potentially result from both the conical distortion and the loss of the gas potential. However comparison with the instantaneous gas removal tests (dotted lines) reveals that the key source of disk heating is actually from the removal of the gas. In most cases, the final stellar disks have an effective thickness roughly equal to that of the instantaneous removal models. This suggests that the removal of the gas potential is the primary cause of disk thickening. 

The conical distortion can be seen to result in an apparent short duration spike of raised effective disk thickness. However this is primarily a result of the way that disk thickeness is measured, assuming a flattened disk (not accounting for a conical distortion), and does not reflect a real spike in disk thickness. However this does confirm the short duration of the conical distortion phase ($<$200 Myr for all models).

In the low concentration, small disk and heavy disk model, the disk thickening is milder with slow winds. In these cases, the effective disk thickness has a value mid-way between the value for instantaneous gas removal and the value of the control model. With slow winds, these disk models maintain a sizeable quantity of gas at inner radii, that continues to provide a source of disk potential. In the outer disk, where the gas has been stripped, the stellar disk expands but in the inner disk it cannot expand.

We conclude that the key source of stellar disk thickening, due to ram pressure stripping, is due to the loss of the gas potential and that the conical distortion has little impact. It is therefore intuitive that the degree of disk heating must depend sensitively on the gas fraction of the original disk. 

This can indeed be seen when comparing the lower gas fraction models (lower two panels) and the standard model (upper left panel): The standard model has a gas to stellar mass ratio of 1 causing the effective disk thickness to increase by $\sim$55$\%$ due to ram pressure stripping. For a gas to stellar mass ratio of 0.3, the disk thickness increases by $\sim 30\%$. For a gas to stellar mass ratio of 0.1, the disk thickness increases by only $\sim 12\%$.

It is difficult to quantify the impact of disk surface density on disk heating from gas removal. In principal a comparison of the standard model with the small and heavy disk model should enable this. The small disk model has a surface density $\sim$70$\%$ larger than the standard model, and yet it too expands by about the same factor ($\sim$55$\%$), suggesting disk surface density is not a strong factor. The heavy disk model initially expands to the value seen in the instantaneous gas loss simulation, but continues to thicken. At this point it has thickened by $\sim55\%$ - equal to the standard model. However the gas disk in this model becomes heavily barred and the rotation of this non-axisymmetric gas distribution results in a secondary source of stellar heating until the bar has been stripped (the final disk is 90$\%$ thicker than the control). 

Disk surface density does not appear to play a key role in determining disk heating unless it causes disk barring. In this current study, we can not draw strong conclusion - as noted previously, we cannot be confident that our simple isothermal treatment of the galaxy gas can realistically treat the complex gas physics involved in bar formation.

Finally, we do see a dependency of disk heating on the halo concentration. Both the standard model (upper left panel) and the low concentration model (upper right panel) have an equal gas-to-stellar mass ratio, yet the disk thickness increases by $\sim 80 \%$ in the low concentration model in comparison to the $\sim55 \%$ in the standard model. This indicates that, while the stellar disks of both models may have gained equal kinetic energy due to the removal of the gas potential, the low concentration model's shallower dark matter well allows the stellar disk to expand to a greater degree.

\section{Discussion}
The current state-of-art in ram pressure stripping modelling typically place strong emphasis on providing the highest possible resolution to the gas disk component of a late-type spiral galaxy. This is indeed necessary to resolve the detailed hydrodynamical mechanisms occurring at the interstellar medium/intercluster medium interface. As a result, such studies (see for example \citealp{Roediger2007}) use analytical potentials to model the gravitational influence of the stellar disk and dark matter halo on the gas dynamics.

In comparison our toy model of ram pressure stripping is hugely idealised and does not contain a live ICM gas component. However, its simplicity has enabled us to treat the dark matter halo and stellar disk with a live component that can dynamically evolve during the ram pressure stripping process. This is clearly crucial to studying the dynamical effects of ram pressure on the stellar disk and dark matter halo component of a galaxy.

The Gunn and Gott condition (see Eqn. \ref{GandGeqn}) has frequently been shown to provide a reasonable first-order predictions of the extent to which a gas disk is truncated by ram pressure. We demonstrate that if the Gunn and Gott condition is satisfied, the force upon the gas disk, and thereby the force transmitted to the stellar disk and surrounding central dark matter, can be analytically calculated. Furthermore, it can be calculated for varying ram pressures, and for disks with a variety of properties.

However, we note one nuance. We have so far assumed that the force on the truncated gas disk is simply the area of the disk (measured to the truncation radius), with the product of the ram pressure (calculated using the product of density of the ICM with the wind velocity squared). This is valid if a galaxy is moving sub-sonically through the ICM. However \cite{Roediger2005} note that this may not be the case once the galaxy moves supersonically through the ICM, thereby generating a shock-wave infront of the disk. While momentum density is conserved perpendicular to a shock front, the velocity $v$ downstream of the shock is reduced while the density is increased. They state that in this situation the ram pressure on the disk behind the shock front is reduced  
by the same factor as $v$. In their simulations, $v$ is reduced by a factor 0.6 and 0.36 for Mach numbers of 1.42 and 2.53 respectively. 

Our toy model does not account for the reduction in ram pressure due to shock fronts from supersonic motion. For the cluster model used in Section \ref{clustermodel}, we note that, for the assumed isothermal gas temperature, the ICM sound speed is $\sim 1000$ km s$^{-1}$. Therefore galaxies with velocities as high as $\sim$2500 km s$^{-1}$ may feel a ram pressure drag force $\sim 65\%$ less strong. However this is less of an issue in the wind-tunnel tests, as we have chosen a fixed cluster ICM density and varied only the galaxy velocity in order to produce a range of ram pressures. For a more dense, but still entirely reasonable choice of ICM wind density, an equal magnitude range of ram pressures can be produced for sub-sonic velocities.

So far we have restricted our study of ram pressure drag effects to a single halo mass of galaxy (10$^{10}$ M$_{\odot}$), perhaps consistent with a medium-mass dwarf galaxy. It is interesting to note that the ram pressure drag force on a giant spiral may be vastly larger. For example, consider a giant bulgeless disk galaxy with a stellar mass of $5 \times 10^{10}$ M$_{\odot}$, a gas fraction of 20$\%$, and a disk scale-length of 2 kpc. Assuming it is moving through an ICM of density equal to that of the wind-tunnel tests with a velocity of 1000 km s$^{-1}$, the drag force upon its stellar and dark matter is a factor $\sim$75 times larger than for our standard model under the same ram pressure. Drag forces are expected to be even stronger if the galaxy contains a significant bulge that increases the restoring force on the gas disk even further. However, the stellar disk mass, and dark matter component that the gas disk must drag is also significantly larger.

Another interesting consideration is the effects of ram pressure drag acting on tidal dwarf galaxies. Such galaxies are believed to contain very little dark matter and be gas rich (\citealp{Duc1998}). If the disks of tidal dwarf galaxies are sufficiently high in surface density to maintain a truncated gas disk against ram pressure, their stellar disks might be heavily effected by ram pressure drag; with little or no dark matter, the stellar disk, alone, would provide the analogous payload on the gas disk. Furthermore, their stellar component should respond far more significantly to the removal of the gas potential than in dark matter dominated models.

Disk thickening is currently the best observational probe of the ram pressure drag effect. However, as disks are only expected to expand by at maximum a factor of two, it may be a challenging effect to observe in dark matter dominated dwarfs. We note that, so far, we have only considered gas fractions as high as 50$\%$ (e.g. the standard and low concentration model). Disk thickening may be stronger in more gas-rich dwarf galaxies, and higher gas fractions become increasingly common at lower galaxy luminosities (\citealp{Gavazzi2008}).

\section{Summary $\&$ Conclusions}
Ram pressure stripping applies a drag force to a truncated gas disk of a spiral galaxy. Gas within the truncated disk cannot be stripped because the self-gravity of the galaxy can overwhelm the ram pressure force. However, this does not mean that the ram pressure drag force simply vanishes.

We demonstrate that from Newton's third law, it is inevitable that the drag force on the gas disk is transmitted to the stellar disk and dark matter halo through their mutual gravitational attraction. Using this approach, we develop a simple analytical equation, based on the Gunn and Gott stripping criteria, that predicts the magnitude of the drag force on disk galaxies of varying properties. 

However the analytical formula cannot predict the response of the stellar disk and dark matter to this forcing. Using a simple toy model of ram pressure, we study the dynamical response of the dark matter and stellar disk to this drag force. We find that the stellar disk, and surrounding central dark matter, can be displaced by ram pressure drag. The displacement is not negligible, and can be of order several kiloparsecs.

We find that our analytical formulation of the drag force can be used to predict the actual drag force to first order for a variety of ram pressures and disk galaxy models. However, it is intrinsically based on the Gunn $\&$ Gott stripping criteria, and therefore may fail if the criteria fails.

Beyond a pure displacement of the stellar and surrounding central dark matter, we find that ram pressure stripping can influence the internal dynamics of the stellar disk. During gas stripping, the stellar disks of our models can develop a slight conical appearance as they are towed along by an inner truncated gas disk. 

This conical phase is brief ($<$200 Myr) and once it is complete, the final stellar disk appears largely unaffected. This indicates that the potential well of the dark matter surrounding the disk is only mildly and temporarily effected by ram pressure drag. 

However the final stellar disks of our models indicate that disk thickening has occurred. We find this is predominantly due to reduction in the total disk potential as a result of the loss of the gas component. The expansion of the stellar disk is therefore strongly dependent on the gas fraction of the original disk. It is also more weakly dependent on the concentration of the halo. For an initially gas rich disk, in a low concentration halo, the stellar disk thickness can increase by close to a factor of two.

Our key results may be summarised as follows. 
\begin{enumerate}
\item The drag force exerted by ram pressure on a truncated gas disk can be transmitted to the stellar disk and surrounding central dark matter of a disk galaxy.
\item This can displace the stellar disk and central dark matter by several kiloparsecs in the direction of the ram pressure wind.
\item The motion of the central dark matter can result in a brief and mild conical-like distortion of the stellar disk, lasting no more than 200 Myr.
\item Disk thickening as a result of ram pressure occurs predominantly due to the reduction in the total disk's potential as a result of gas loss. Disks can thicken by as much as a factor of two as a result of ram pressure in disks with a gas fraction of 50$\%$.
\end{enumerate}

\section*{Acknowledgements}
RS is funded through ESO's Comite Mixto. MF announces financial support from FONDECYT grant 1095092 and the Chilean Centro de Excelencia en Astrofisica y Tecnologies Afines (CATA).
\bibliography{bibfile}

\bsp

\label{lastpage}

\end{document}